\documentclass[10pt,twocolumn,letterpaper]{article}

\usepackage[pagenumbers]{cvpr} % To force page numbers, e.g. for an arXiv version
\usepackage[dvipsnames]{xcolor}

\definecolor{cvprblue}{rgb}{0.21,0.49,0.74}
\usepackage[pagebackref,breaklinks,colorlinks,citecolor=cvprblue]{hyperref}

\def\confYear{2024}

\title{Improving Multimodal Learning with Multi-Loss Gradient Modulation}

\author{Konstantinos Kontras\\
ESAT, KU Leuven\\
Leuven, Belgium\\
{\tt\small konstantinos.kontras@kuleuven.be}
\and
Christos Chatzichristos\\
ESAT, KU Leuven\\
Leuven, Belgium\\
{\tt\small christos.chatzichristos@kuleuven.be}
\and
Matthew Blaschko\\
ESAT, KU Leuven\\
Leuven, Belgium\\
{\tt\small matthew.blaschko@kuleuven.be}
\and
Maarten De Vos\\
ESAT, KU Leuven\\
Leuven, Belgium\\
{\tt\small maarten.devos@kuleuven.be}}
\usepackage{multirow}
\usepackage{arydshln}
\usepackage{graphicx}
\usepackage{adjustbox}
\usepackage{cuted}
\usepackage{etoolbox}
\AfterEndEnvironment{strip}{\leavevmode}

\begin{document}
\maketitle
\begin{abstract}
Learning from multiple modalities, such as audio and video, offers opportunities for leveraging complementary information, enhancing robustness, and improving contextual understanding and performance. However, combining such modalities presents challenges, especially when modalities differ in data structure, predictive contribution, and the complexity of their learning processes. It has been observed that one modality can potentially dominate the learning process, hindering the effective utilization of information from other modalities and leading to sub-optimal model performance. To address this issue the vast majority of previous works suggest to assess the unimodal contributions and dynamically adjust the training to equalize them. We improve upon previous work by introducing a multi-loss objective and further refining the balancing process, allowing it to dynamically adjust the learning pace of each modality in both directions, acceleration and deceleration, with the ability to phase out balancing effects upon convergence. We achieve superior results across three audio-video datasets: on CREMA-D, models with ResNet backbone encoders surpass the previous best by 1.9\% to 12.4\%, and Conformer backbone models deliver improvements ranging from 2.8\% to 14.1\% across different fusion methods. On AVE, improvements range from 2.7\% to 7.7\%, while on UCF101, gains reach up to 6.1\%.
\end{abstract}    
\section{Introduction}

Combining data from several modalities such as vision, text, audio, and time series has significantly improved performance across many tasks and has proven particularly advantageous in cases of noisy or unreliable sources \cite{Beit, li2022blip, kontras2024core, radevski2021revisiting, radevski2023multimodal}. However, studies show that the inclusion of a new modality doesn't always benefit, and can even impair, model performance \cite{wang2020makes}. As explained by Huang \etal \cite{huang2022modality}, different modalities compete with each other, resulting in underperforming modalities. These modalities exhibit inferior performance in their multimodal-trained encoders compared to their unimodal counterparts, assessing each modality encoder independently post-training, suggesting potential imbalances or inefficiencies in the integrated training process. This finding contradicts the assumption that more information necessarily improves task understanding.

To mitigate this effect, various balancing strategies have been proposed \cite{OGMGE, AGM, PMR, MSLR, fujimori2020modality, wang2020makes, wu2022characterizing, xsleepnet}. Previous methods typically use models with individual unimodal encoders and a fusion network that produces the multimodal output, generally falling into two main categories. Methods in the first category adjust the learning rate of unimodal encoders based on their estimated predictive performance \cite{OGMGE, MSLR, wu2022characterizing}. Meanwhile, the methods of the second category employ additional loss functions derived from unimodal predictions and balance these losses during multimodal learning \cite{vielzeuf2018centralnet, wang2020makes, xsleepnet, PMR, du2023uni}. These losses enable accurate estimation of unimodal performance, however, these methods do not address the effects caused by the multimodal loss. We aim to bridge both approaches by incorporating additional losses and simultaneously adjusting the learning rates. Our methodology provides the following improvements and novelties:

\begin{figure*}[t]
% \hspace{-5mm}
\includegraphics[width=\linewidth]{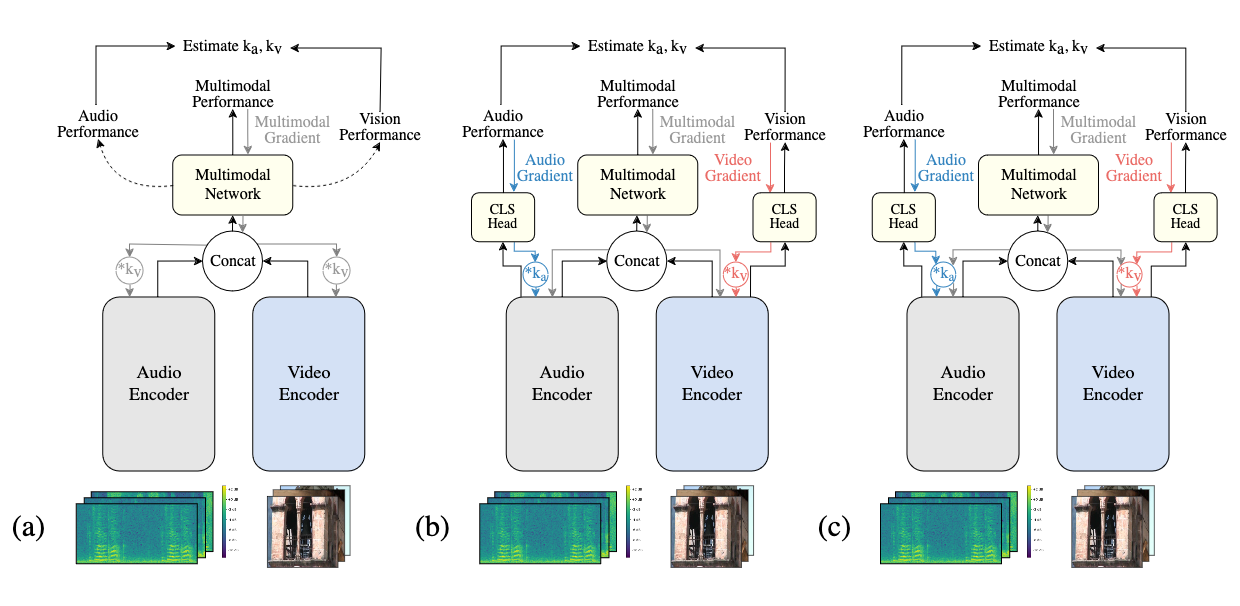}
\caption{
Categorization of state-of-the-art balancing methodology: (a) Gradient Balancing methods use estimates of unimodal performance to calculate coefficients ($k_a$ and $k_v$) and employ these to balance the multimodal loss. (b) Multi-Task methods incorporate unimodal classifiers into the model, each noted as CLS Head, to better estimate unimodal performance. The coefficients ($k_a$ and $k_v$) derived from comparing unimodal performance are used exclusively to balance the unimodal losses. (c) The proposed Multi-Loss Balanced method combines both strategies by incorporating unimodal classifiers for accurate unimodal performance estimation and balancing both multimodal and unimodal losses.
}
\label{fig:methods}
\end{figure*}

\begin{itemize}
\setlength\itemsep{0mm}
\item Throughout the entire training process, we employ a multi-objective loss that not only ensures each encoder converges close to its unimodal optimum, similarly to the methods described in \cite{vielzeuf2018centralnet} and \cite{wang2020makes}, but also facilitates accurate unimodal performance estimation. We apply balancing for both multimodal and unimodal losses, distinguishing our method from any prior efforts in these categories.
\item The proposed balancing technique, inspired by \cite{OGMGE}, adaptively modifies the learning rates of encoders based on unimodal performance assessments. We enhance this strategy by enabling both acceleration and deceleration, tailored to the relative performance of the modalities.
\item The balancing equations are designed such that when all unimodal encoders converge, the balancing naturally phases out, eliminating the need to pre-determine explicitly which epoch should mark the end of balancing.
\end{itemize}
Results conclusively show that the suggested method consistently outperforms balancing state-of-the-art methods across three video-audio datasets, employing either ResNet or Conformer backbone encoder models, and utilizing a range of fusion techniques.
% 

% \begin{figure}
%     \centering
%     \includegraphics{Figures/Methods_Workshop_1GD.drawio.pdf}
%     \caption{Caption}
%     \label{fig:enter-label}
% \end{figure}

% \input{sec/2_modality_imbalance}    

\section{State-of-the-Art}
\label{sec:prev_works}

Previous methods addressing the challenge of modality competition in multimodal learning frameworks can mostly be broadly categorized into two primary groups: Gradient Balancing and Multi-Task methods. Those are illustrated in (a) and (b) of Figure \ref{fig:methods}. Additionally, methods that don't fit into either category are %considered supplementary and 
described in Section~\ref{sec:OtherApproaches}.

\subsection{Gradient Balancing Techniques}

In this category, methods address modality competition by adjusting the unimodal encoders' learning rates. The Modality-specific Learning Rates (\textbf{MSLR})  \cite{MSLR} strategy introduces an approach for decision-level fusion models which dynamically adjust encoder gradients magnitude based on recent validation accuracy, independently for each modality. Building on this foundation, On-the-fly Gradient Modulation (\textbf{OGM}) \cite{OGMGE} introduces an interactive framework, comparing performance improvements across modalities to tailor the adjustment of each encoders' learning rate. These methods are designed for late fusion models, allowing direct access to unimodal performance; however, they cannot directly be applied in more complex fusion schemes. Wu et al. \cite{wu2022characterizing} partially belong to this category, as they propose a method that monitors the learning speed of each modality through gradient norms while using isolated phases of unimodal training to balance multimodal learning. However, measuring this learning speed requires a specialized model architecture, which prevents the use of a shared multimodal output, limiting its application.

\subsection{Multi-task Learning}

The methods that belong in this category target the sub-optimal unimodal encoders by employing dedicated losses for each modality. It has been observed that incorporating unimodal classifiers enhances multimodal learning, as demonstrated in \cite{vielzeuf2018centralnet, kontras2024core}. Building on this foundation, Wang \etal \ \cite{wang2020makes} suggest to dynamically adjust the weights of the unimodal losses based on an overfitting-to-generalization ratio, however accessing that information requires a separate validation set. Du \etal \ \cite{du2023uni} introduce a teacher-student schema with distillation losses for direct guidance of unimodal encoders. The Prototypical Modal Rebalance (\textbf{PMR}) method \cite{PMR} circumvents the use of additional parameters for classifiers by leveraging prototype-based classifiers and distance-based losses. This approach maintains a similar logic, wherein imbalanced performance is compensated for by the appropriate unimodal loss. However, balancing the unimodal losses alone disregards the effects caused by the multimodal loss.

\subsection{Other Approaches}%Supplementary Methods}
\label{sec:OtherApproaches}

Additional studies explore further approaches to multimodal learning, not fitting neatly into previously discussed categories. \textbf{MMCosine} \cite{xu2023mmcosine} preconditions late fusion by standardizing both the feature vectors and the weights dedicated to each modality, equalizing their contribution to the final prediction. Gat \etal \cite{gat2020removing} explore a regularization technique to enhance each modality's contribution by maximizing functional entropy, estimated through prediction differences after input perturbations. This method, however, increases the model's sensitivity to those perturbations. Adaptive Gradient Modulation (\textbf{AGM}) \cite{AGM} takes the contribution of each modality by utilizing zero-masking Shapley \cite{shapley1953value, lundberg2017unified} values. This allows for balancing on models of any structure, however at the cost of increased computational demands due to multiple forward passes.  

\section{Method}

\subsection{Model}

We have endeavored to maintain our methodology as model-agnostic as possible, albeit with a few necessary assumptions. Our focus centers on multimodal models, wherein each modality is processed by a dedicated encoder. We aim to enhance the training efficiency of these encoders, thereby indirectly benefiting the entire network's performance. To elucidate, our models conform to a unified structural framework.

Given a dataset $D$ consisting of $N$ samples across $M$ modalities, denoted as $X = \{X_1, .., X_M\}$, each sharing a common ground truth label $Y$ with $C$ distinct classes, we employ networks structured as follows:

\begin{equation}
    f(X;\theta) = f_{v}(f_1(X_1;\theta_1), .., f_M(X_M;\theta_M) ; \theta_{v}),
\end{equation}

\noindent where $\theta$ denotes all the parameters of the network, $\theta_1, .., \theta_M$ the unimodal encoders', $f_1, .., f_M$ parameters and $\theta_{v}$ the common parameters of the common fusion function $f_{v}$.

\subsection{Balancing}

Our objective is to achieve synchronous convergence of the unimodal encoders, based on the hypothesis that this will prevent the model from overfitting to any single modality and may encourage the development of more synergistic behaviors. 

To achieve this, following prior studies \cite{PMR, wang2020makes, OGMGE, MSLR}, we dynamically adjust the learning rates of each unimodal encoder based on the comparative analysis of unimodal performances. We estimate balancing coefficients for each modality $i=1, .., M$ as $k_i$ that indicate how much each modality should change the learning pace of its encoder. When $k_i > 1$, we accelerate the learning of the modality, while when $k_i < 1$, we decelerate it. The update rule using stochastic gradient descent (SGD), the initial learning rate $lr_{\text{base}}$ and the loss function $L$ would then be:

\begin{equation}
    \Delta \theta_i = - lr_{\text{base}} \cdot k_i \cdot \nabla L(X,y,\theta).
\label{eq:update}
\end{equation}

We improve upon previous works in three points. Firstly, we incorporate a multi-task objective with additional unimodal losses. Such objective ensures precise assessment of unimodal performance, while aiding the convergence of the unimodal classifiers. Secondly, our method allows for both the acceleration and deceleration of learning across the modalities, without ever entirely halting their progress. Finally, while previous methods imposed a hard limit to cease balancing after a predetermined number of epochs, we employ a function that naturally reduces balancing effects as the performances of both modalities converge.

Our multi-loss objective $L$ is a summation of a cross-entropy ($CE$) loss for the multimodal predictions and similar cross-entropy losses for each unimodal prediction. This can be expressed as:

\begin{equation}
    L = CE(f(X;\theta), y) + \sum^M_{i=1}{CE(f_i(X_i;\theta_{i}), y)}.
\end{equation}

% To adaptively balance the learning of each unimodal encoder, we derive the balancing coefficients $k_i$ of Eq. \ref{eq:update} based on each modality performance. These coefficients are calculated as follows:\

To adaptively balance the learning of each unimodal encoder, we derive the coefficients $k_i$ of Eq. \ref{eq:update} based on modality performance. These coefficients are calculated as follows:

\begin{align}
    s_i &= \sum_{i=1}^M \sum_{c=1}^Cf_i(X_i;\theta_{i})1_{k=y_c}, \\ 
    r_i &= \frac{\frac{1}{M-1}\sum_{j=1, j\neq i}^Ms_j}{s_i}, \label{eq:ri}\\
    \beta_i &= \begin{cases} 
    \begin{aligned}
& \beta_{max} \quad \text{if } r_i > 1,\\
& 2 \quad \text{otherwise},
\end{aligned}
\end{cases} \\
    k_i &= 1 + (\beta_i-1) \cdot \tanh(\alpha \cdot (r_i - 1)) \label{eq:ki}
    % k_i &= 1+\tanh(\alpha \cdot (r_i -1)), \label{eq:ki_norm}
\end{align}

\noindent $s_i$ is the sum of correct class predictions by the unimodal encoder for modality $i$, $r_i$ the relative performance of each encoder compared to the average performance of the others, $\beta_i$ the max value of the coefficients and $k_i$ the final balancing coefficient for each modality, where $\alpha \in \mathbb{R}^+$ and $\beta_{max} \in \mathbb{R}^+$, $\beta_{max} \geq 1$ are predetermined hyperparameters. Finally, $tanh$ is the hyperbolic tangent function, providing a smooth normalization of $r_{i}$. 

% The use of two distinct cases on $b_i$ arise from the different behaviors we want our balancing method to perform. When $r_i <1 $ it means that other modalities are performing on average worse than modality $i$ and we want to reduce the learning rate of modality $i$ to slow down its learning pace. To achieve that we choose $b_i=2$ and we end up in the part of Figure \ref{fig:kivsri} with $r_i \in (0,1)$. The range of the learning slowing down can be of any magnitude and can be already affected by the slope of the $tanh$ with parameter $a$, greater $a$ will lead to more radical changes in the learning rate. On the other hand when $r_i>1$ meaning that the other modalities on average are performing better, we want to increase the learning rate of modality $i$ and therefore we need $k_i>1$. The magnitude of the increase is more sensitive in the acceleration, since large numbers will lead the model to diverge. In different models we expect the potential increase to differ and therefore we set the maximum increase as the hyperparameter $b_{max}$. For example, if  $b_{max}=10$, we allow the learning rate to increase up to 10 times depending on the performance ratio $r_i$. This can be seen in Figure \ref{fig:kivsri} when $r_i >1 $

\begin{figure}[ht]
\centering
\includegraphics[width=0.9\linewidth]{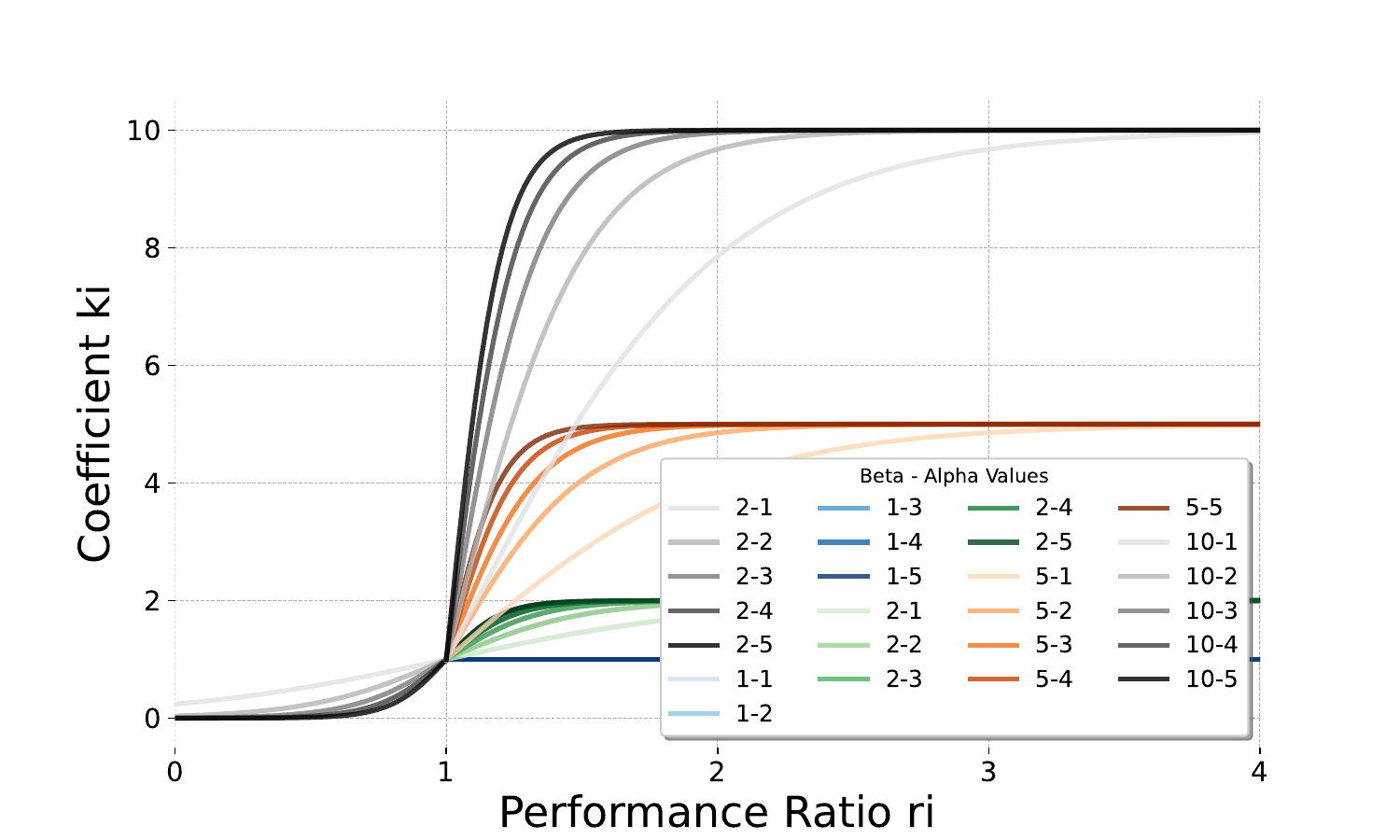}
% \caption{Balancing Coefficients $k_i$ Eq. \ref{eq:ki} vs Performance Ratio $r_i$ Eq. \ref{eq:ri} with multiple $\alpha$ and $\beta$ values.}
\caption{Comparing Balancing Coefficients ($k_i$, Eq. \ref{eq:ki}) and Performance Ratios ($r_i$, Eq. \ref{eq:ri}) across different $\alpha$ and $\beta$ settings.}
\label{fig:kivsri}
\end{figure}

The utilization of two distinct cases for $b_i$ arises from the varied behaviors we aim to achieve with the balancing method. When $r_i<1$, it indicates that other modalities perform, on average, worse than modality $i$, necessitating a reduction in the learning rate of modality $i$ to slow its learning pace. To achieve this, we set $b_i=2$, placing us in the portion of Figure \ref{fig:kivsri} where $r_i \in [0, 1]$. The extent of the learning deceleration can vary and is influenced by the slope of the $tanh$ function parameterized by $a$; a larger $a$ value results in more pronounced changes in the learning rate. Conversely, when $r_i>1$, indicating that other modalities generally outperform modality $i$, we aim to increase the learning rate of modality $i$, requiring $k_i>1$. The extent of the acceleration is more sensitive, as large values may lead to model divergence. To control this, we introduce the hyperparameter $b_{max}$, which sets the maximum permissible increase. For instance, if $b_{max}=10$, we allow the learning rate to increase by up to 10 times based on the performance ratio $r_i$, as shown in Figure \ref{fig:kivsri} when $r_i >1$.

\section{Experiments}

\subsection{Datasets}
\label{sec:datasets}

\textbf{CREMA-D \cite{cao2014crema}:} is an emotion recognition dataset with audio and video modalities, featuring 91 actors expressing six emotions. Video frames are sampled at 1 frame-per-second (fps), selecting 3 consecutive frames, while audio segments are sampled at 22 kHz. Audio analysis employs Short-Time Fourier Transform (STFT) with a window size of 512 and a step size of 353 samples to create the log-Mel spectrograms. For advanced models, we adopt preprocessing steps suggested by Goncalves \etal \cite{Goncalves_2023_2}, utilizing the full audio recording at 16kHz without STFT and video data without subsampling. We also follow their dataset division, ensuring no actor overlap between training, validation, and test sets. Standard deviation (std) is reported across 3-folds.

\textbf{AVE \cite{tian2018audio}:} spans 28 event categories of everyday human and animal activities, each with temporally labeled audio-video events lasting at least 2 seconds. Video segments with the event are sampled at 1 fps for 4 frames, with audio resampled at 16 kHz using CREMA-D's STFT settings. AVE provides predefined training, validation, and test splits. Standard deviation (std) is calculated from three random seeds on the same test set.

\textbf{UCF101 \cite{soomro2012ucf101}:} showcases real-life action YouTube videos across 101 categories, expanding UCF50. Our analysis focuses on 51 action categories featuring both video and audio modalities, following similar data preparation as the AVE dataset. Model evaluation utilizes the dataset's 3-fold split, with std reported across these folds.

\setlength{\fboxrule}{0pt}
\begin{figure*}[t]
    \centering
    \begin{subfigure}[t]{\textwidth}
        \centering
        \hspace{18mm}
        \includegraphics[width=0.8\textwidth]{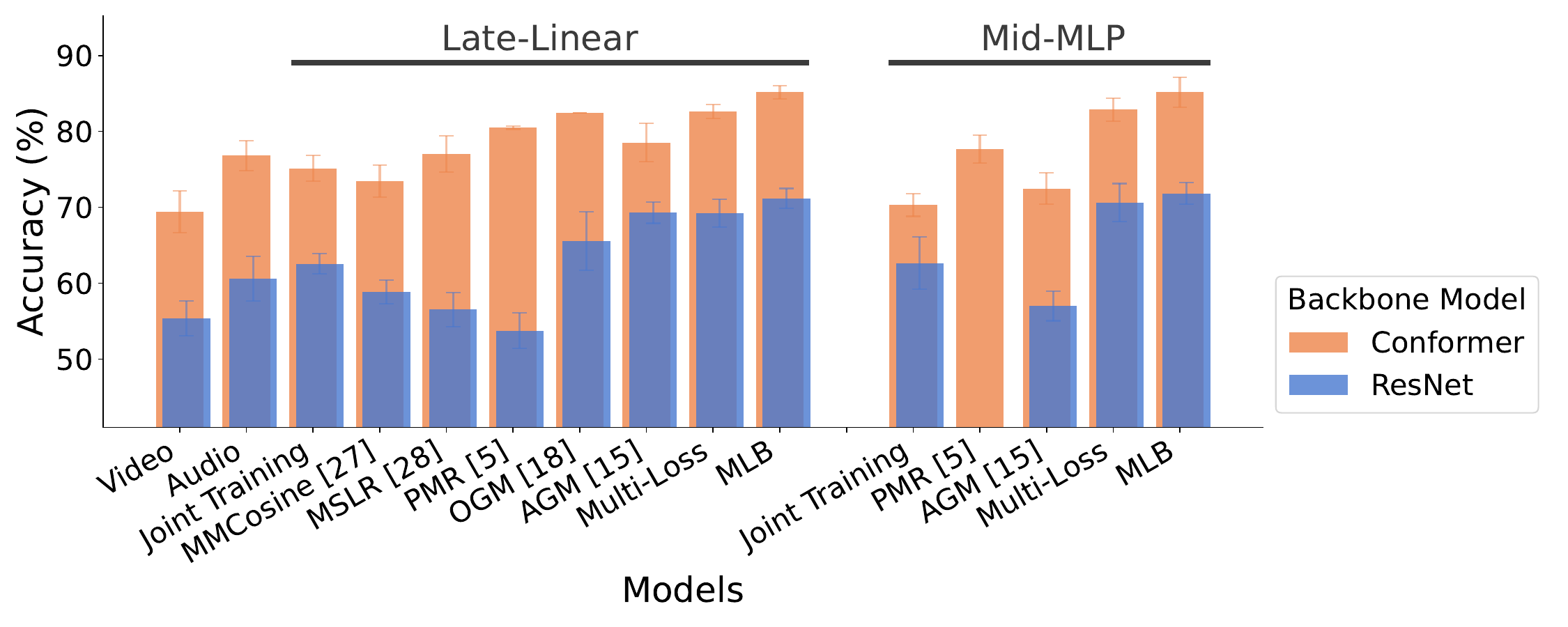}
        \caption{CREMA-D}
    \end{subfigure}
    \begin{subfigure}[t]{0.47\textwidth}
        \centering
        \includegraphics[width=\textwidth]{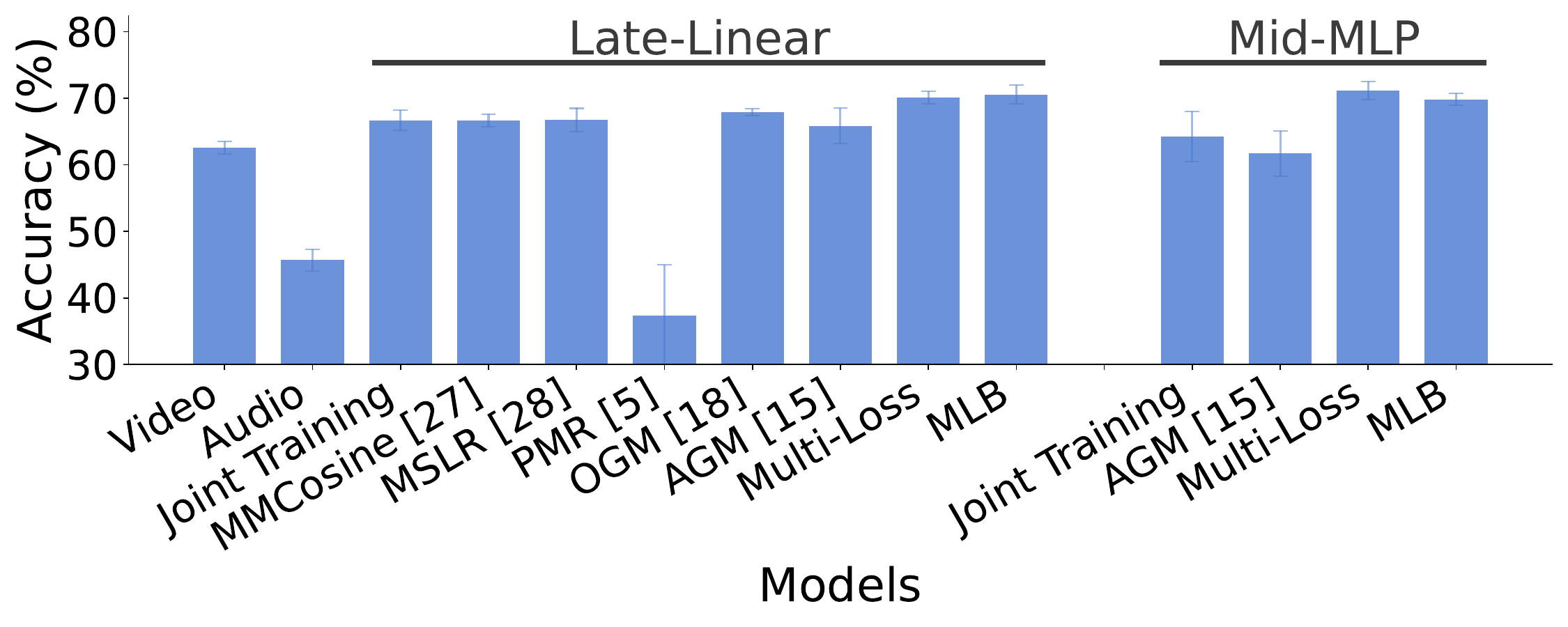}
        \caption{AVE}
    \end{subfigure}
    \begin{subfigure}[t]{0.47\textwidth}
        \centering
        \includegraphics[width=\textwidth]{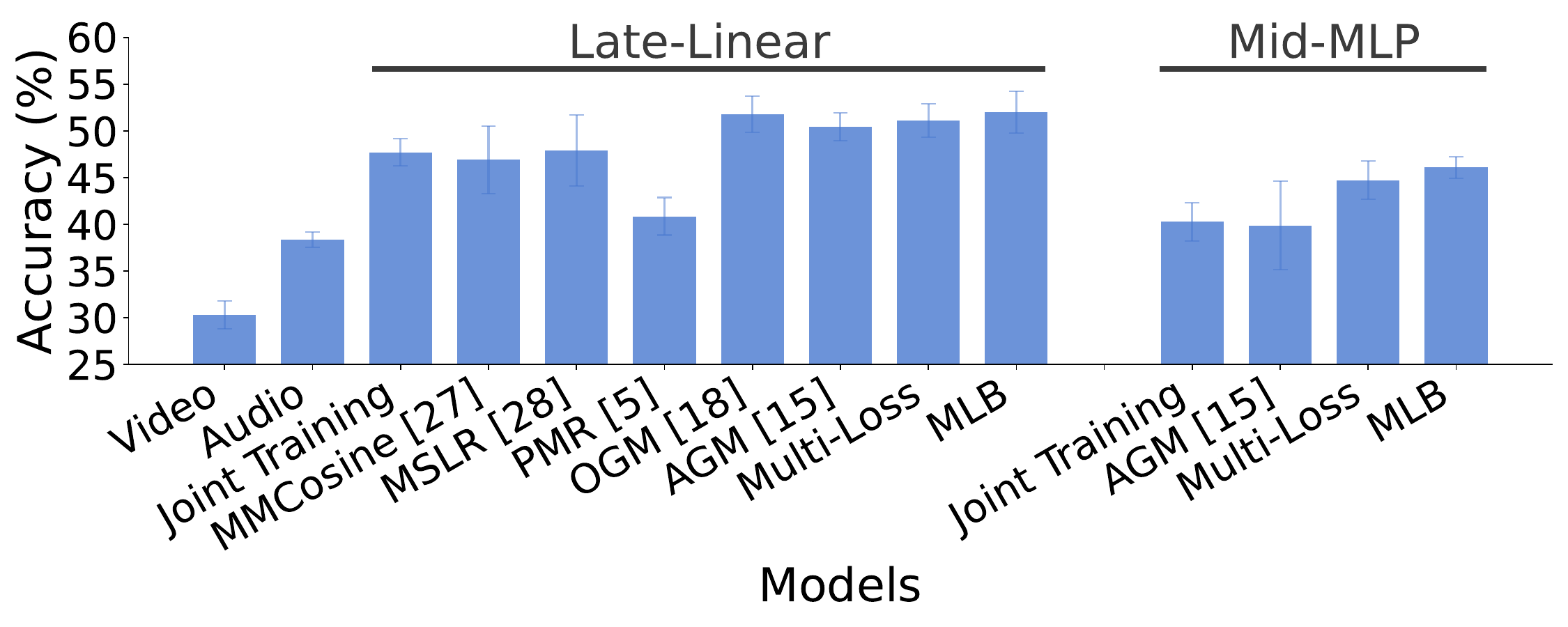}
        \caption{UCF-101}
    \end{subfigure}
    % \begin{subfigure}
    %     \includegraphics[width=0.8\linewidth]{Figures/bar_results_CREMA-D.pdf}
    %     \caption{(a) CREMA-D}
    % \end{subfigure}
    % \begin{subfigure}
    %     \includegraphics[width=0.5\linewidth]{Figures/bar_results_AVE.pdf}
    %     \caption{(b) AVE}
    % \end{subfigure}
    %     \begin{subfigure}
    %     \includegraphics[width=0.5\linewidth]{Figures/bar_results_UCF101.pdf}
    %     \caption{(c) UCF-101}
    % \end{subfigure}

        % \hspace*{15mm}\includegraphics[width=0.8\linewidth]{Figures/bar_results_CREMA-D.pdf}
        % \fbox{\parbox{0.94\textwidth}{\centering (a) CREMA-D}}
    %     \includegraphics[width=0.47\linewidth]{Figures/bar_results_AVE.pdf}
    %     \includegraphics[width=0.47\linewidth]{Figures/bar_results_UCF101.pdf}
    % \fbox{\parbox{0.47\textwidth}{\centering (b) AVE}}
    % \fbox{\parbox{0.47\textwidth}{\centering (c) UCF-101}}
    \caption{Accuracy of models that differentiate by the backbone encoders (colors), the fusion strategies (Late with a linear and Mid with a MLP classifier), and the balancing techniques (x-axis). Across all datasets, results demonstrate that employing unimodal losses within the Multi-Loss framework and balancing them consistently yields the best performance.}
    \label{fig:perf_results}
\end{figure*}

\subsection{Backbone Unimodal Encoders}

Our experimental framework aims to demonstrate the broad applicability of our findings across diverse unimodal encoders. We used two type of encoders for each modality: one randomly initialized ResNet and a Transformer-based with pretrained weights.

In line with prior research \cite{AGM, OGMGE, PMR, xu2023mmcosine}, we employed ResNet-18 \cite{he2015deep}, initialized from scratch, as the unimodal encoder for handling both video and audio modalities across all datasets. We extend our analysis to include larger, pre-trained models. Following \cite{goncalves2023versatile} on CREMA-D, we deploy the first 12 layers of the Wav2Vec2 \cite{baevski2020wav2vec, wolf-etal-2020-transformers} model with self-supervised pretrained weights for speech recognition. For the video modality we extract the facing bounding boxes and afterwards the facial features exploiting EfficientNet-B2 \cite{tan2019efficientnet} as a frozen feature descriptor. Both the audio and video features are further refined using each a 5-layer Conformer \cite{gulati2020conformer}. Leveraging similar audio pre-trained encoders for the AVE and UCF-101 datasets did not yield significant improvement due to differences in data distribution, as these datasets lack speech-based audio. As a result, we focused exclusively on experimenting with ResNet for these two datasets.

\subsection{Multimodal Fusion}
\label{sec:fusion_methods}

% Having established the unimodal backbone encoders, our next step is to determine the type of multimodal fusion we will explore. Our investigation will focus on models that eventually integrate all modalities at some point within the network.
We are outlining five fusion models that combine unimodal features from the encoders to generate the multimodal output. Late Fusion combines the concatenated unimodal features using a linear layer, noted as \textbf{Late-Linear}. Mid Fusion employs a 2-layer Multilayer Perceptron (MLP) to investigate the effects of nonlinear feature combinations on modality imbalances, noted as \textbf{Mid-MLP}. We align with prior research and experiment with advanced fusion strategies: Feature-wise Linear Modulation (\textbf{FiLM})\cite{perez2018film} and \textbf{Gated} mechanisms \cite{kiela2018efficient}. Additionally, we explore integrating a 2-layer Conformer \cite{gulati2020conformer} model as a Transformer-based (\textbf{TF}) fusion method, utilizing both class and modality-independent tokens. Our objective is to showcase the impact of balancing methods by assessing their effectiveness across diverse fusion strategies.

% We conducted a grid search to tune hyperparameters for each balancing method, ensuring fairness with an equal number of trials. Different datasets, backbone encoders, and fusion strategies underwent separate hyperparameter searches. Our findings indicated stable hyperparameter values across methods, but tuning was essential for each one to reveal the benefit of each method. The complete list of hyperparameters is available in Appendix \ref{supp: hyperparameters}.

\section{Results}
\label{section:results}

In this section, we present results on the effectiveness of using different fusion strategies on multimodal training with several multimodal balancing techniques namely MMCosine, MSLR, OGM, AGM, PMR\footnote{The PMR method, as per the provided code, exhibits several instabilities, resulting in exploding values of the classifier prototypes. This instability predominantly affects models with ResNet encoders and certain nonlinear fusion methods. Despite these issues, we opt not to modify their implementation to avoid inadvertently influencing the method incorrectly.}, Multi-Loss and Multi-Loss Balanced (\textbf{MLB}). We show the models trained under the single multimodal objective without any balancing methods, named as \textbf{Joint Training}. Hyperparameter tuning was conducted separately for each method, dataset, backbone encoder, and fusion strategy on the validation set. This ensured fairness with an equal number of trials for each configuration.

\setlength{\fboxrule}{0pt}
\begin{figure*}[t]
    \centering
    % \bgroup
    % \def\arraystretch{0.9}
    % \begin{tabular}{c}
        \includegraphics[width=0.8\linewidth]{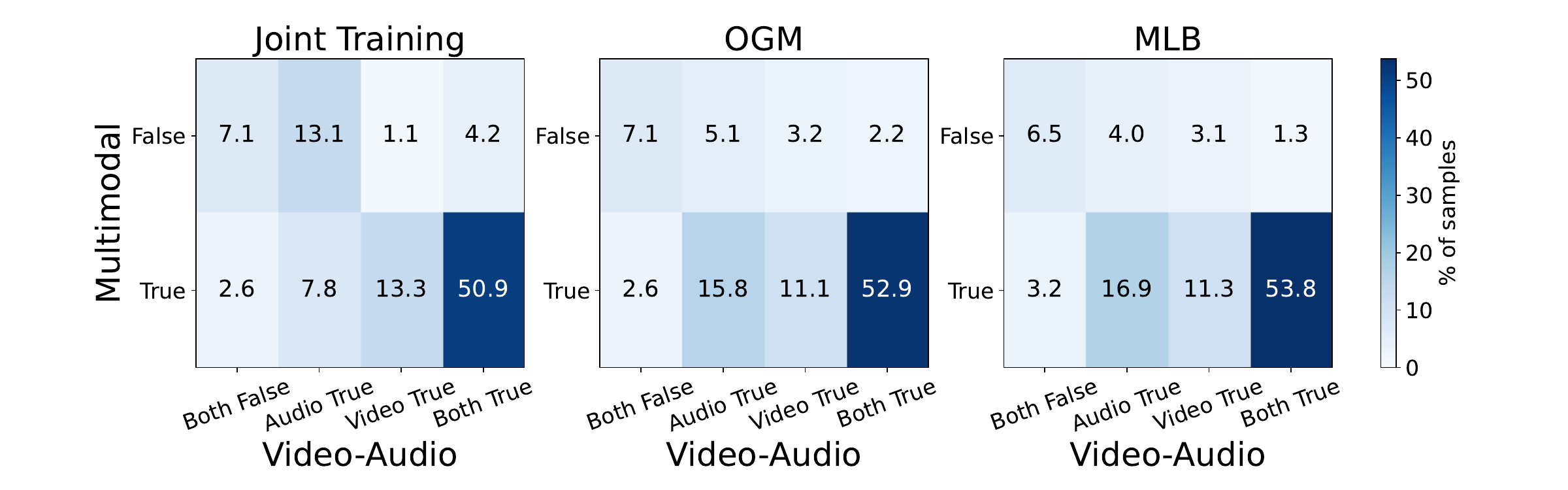}
        \fbox{\parbox{\textwidth}{\centering (a) Conformer}}
        \includegraphics[width=0.8\linewidth]{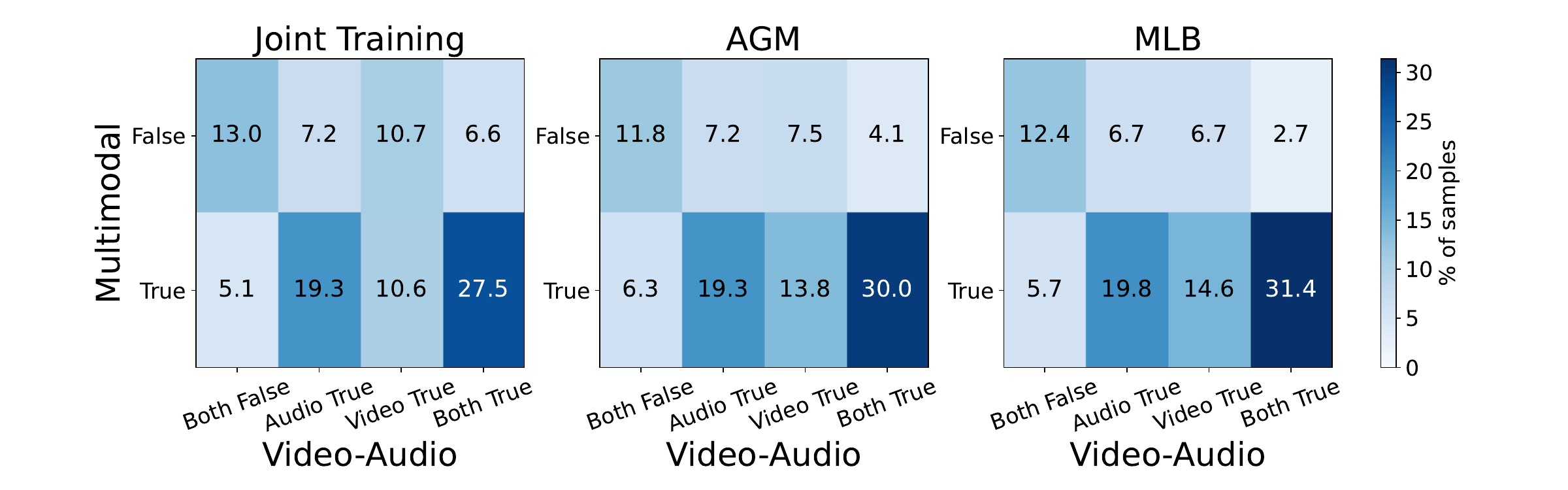}   \fbox{\parbox{\textwidth}{\centering (b) ResNet}}
    % \end{tabular}
    % \egroup

    \caption{Confusion Matrices between the unimodal models and the multimodal ones for both backbone encoders, (a) Conformer and (b) ResNet, trained under different balancing methods. Each column of the confusion matrix represents the cases where both unimodal predictions are incorrect, where only one is correct, and where both are correct. MLB consistently balances and improves performance across all categories.}

\label{fig:confusion matrices}
\end{figure*}

In the bar plots presented in Figure \ref{fig:perf_results}, we assess the previously mentioned models alongside suitable balancing methods across the three datasets. Results highlight that MLB consistently outperforms previous approaches, demonstrating its robustness across various settings. Our comparison with the simpler Multi-Loss strategy reveals that adding unimodal losses often suffices to adequately train the unimodal encoders. In most instances balancing provides a further incremental benefit.

\begin{figure}[h]
    \centering
    \includegraphics[width=0.9\linewidth]{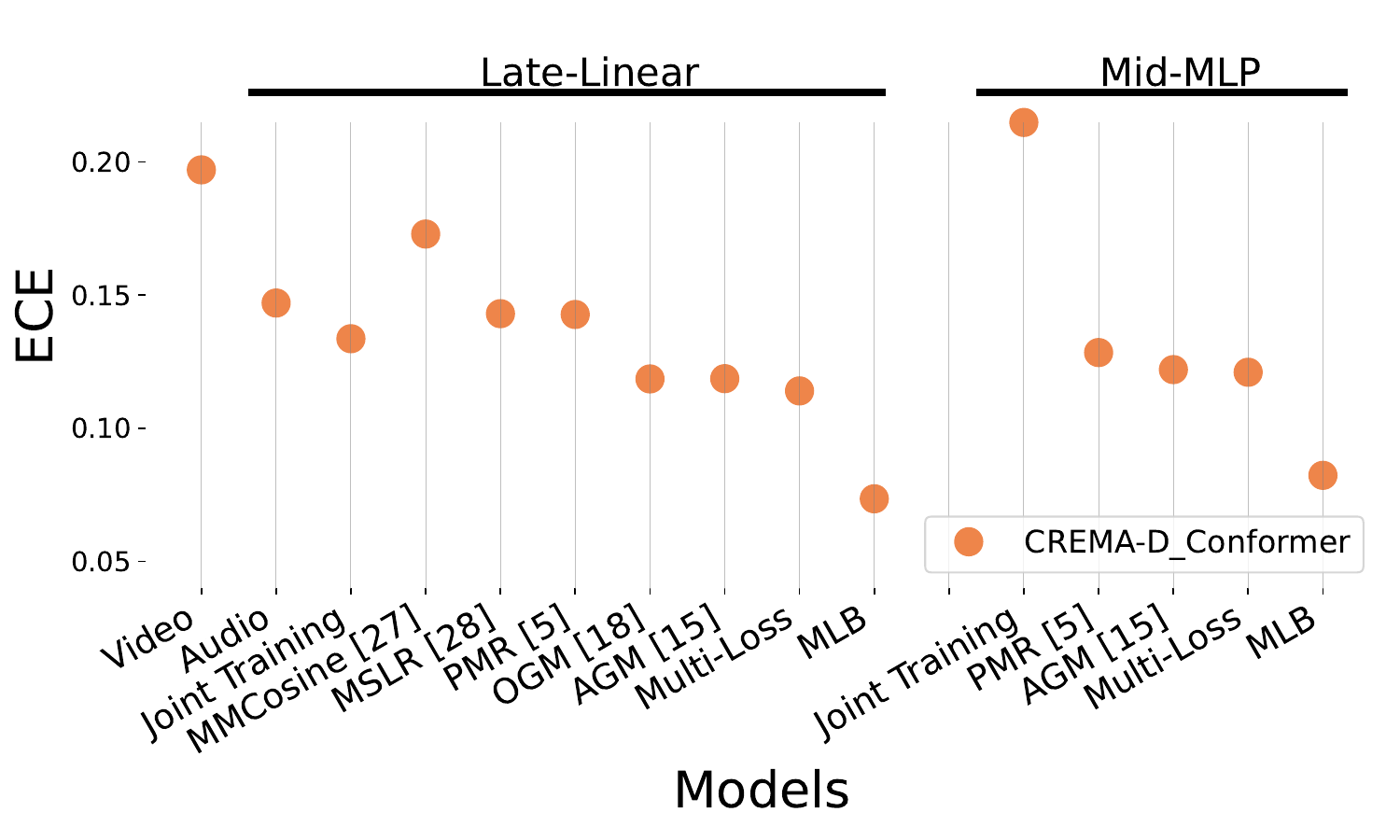}
    \caption{ECE Comparison on Conformer CREMA-D.}
    \label{fig:ece}
\end{figure}

Furthermore, we use the Expected Calibration Error (ECE) \cite{guo2017calibration} to assess whether multimodal training enhances model uncertainty awareness. Figure \ref{fig:ece} demonstrates that the MLB method, achieving the highest accuracy, also exhibits lower calibration error compared to other methods and to Multi-loss, highlighting the significance of balancing.

\begin{figure}[h]
    \centering
    \hspace*{15mm}\includegraphics[width=\linewidth]{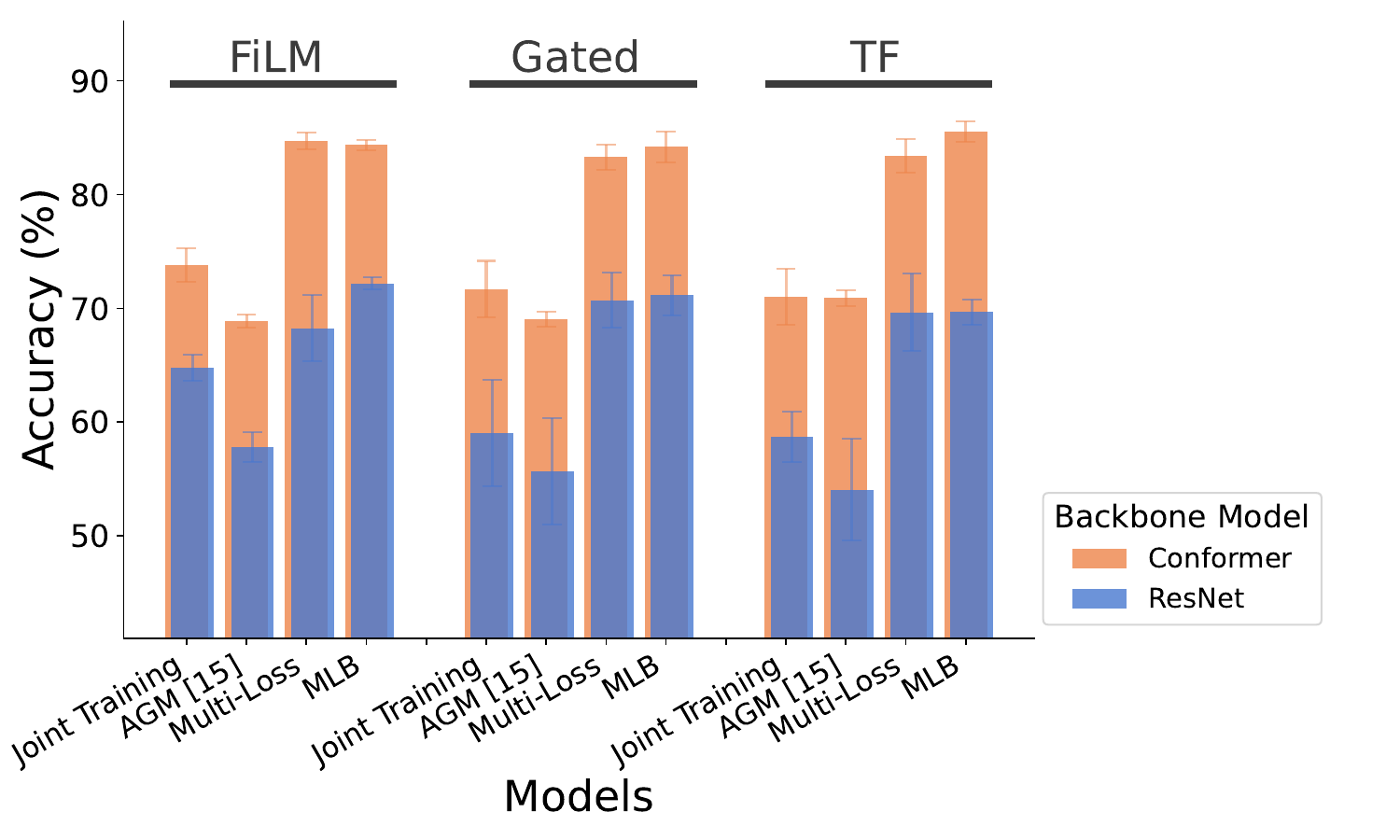}

    \caption{Accuracy of models using FiLM \cite{perez2018film}, Gated \cite{kiela2018efficient} and TF fusion techniques on the ResNet and Conformer model applied to the CREMA-D dataset.}
    \label{fig:fusiongates}
\end{figure}

To verify that the issue of overfitting on one of the modalities is not prevalent to the fusion method, we incorporate alternative fusion strategies, as mentioned in Section \ref{sec:fusion_methods}, FiLM, Gated and TF. Both the Multi-Loss and Multi-Loss Balanced employ additional linear classifiers to derive unimodal predictions and calculate the unimodal losses. Figure \ref{fig:fusiongates} supports our primary findings, indicating improvements in the learning process across models employing different fusion methods balanced with Multi-Loss and MLB. However, in some instances, we do not observe notable benefits from the addition of the balancing technique.

To ascertain whether the lack of training stems from an underperforming modality in the Joint Training schema and to assess the impact of different balancing techniques, we examine the confusion matrices. In Figure \ref{fig:confusion matrices}, we illustrate the label agreement among the two unimodal models and the multimodal ones trained under various methods. Notably, Joint Training with the Conformer model tends to overfit to the video modality, exhibiting high agreement with it. Conversely, the same method applied to the ResNet model demonstrates a stronger reliance on the audio modality, resulting in increased errors on the Video True case. Interestingly, both the pretrained Conformer for video and the ResNet for audio required fewer optimization steps to converge compared to the other modality. This observation leads us to conclude that a modality's dominance in the learning process is not solely determined by its predictive power, but also by its ability to reduce the training loss more rapidly.

From the same figure, MLB consistently enhances performance across all categories, including Both True, Both False, Audio True, and Video True, indicating across-the-board improvement. In the ResNet model results, we observe that AGM, the best previous method, slightly enhances the scenario where both modalities predict incorrectly (Both False), suggesting improved information discovery.

\section{Discussion and Conclusion}

In this paper, we tackle the issue of multimodal models overfitting on one modality, thereby hindering the effective utilization of the remaining modalities. We experiment with state-of-the-art balancing methods, revealing inconsistencies across different models and datasets. To address this, we introduce the Multi-Loss Balanced method. We demonstrate how additional losses derived from each modality can enhance the training of unimodal encoders and provide accurate estimations of their performance, enabling us to balance the learning of each encoder during multimodal training. Our approach incorporates coefficient estimation functions to support both acceleration and deceleration of each modality, while allowing the model to mute the balancing as it converges. We consistently observe improved results across three video-audio datasets, utilizing both ResNet and Transformer-based backbone encoders, along with a variety of fusion methods.
% \begin{equation}
%     \min_{\theta_1}\max_{\theta_0} dist( f_{\theta}(X_0, X_1), f_{\theta}(P_0, X_1) )^3
% \end{equation}

% \begin{equation}
%     \min_{\theta_0}\max_{\theta_1} dist( f_{\theta}(X_0, X_1), f_{\theta}(X_0, P_1) )^3
% \end{equation}

% \begin{align*}
%     dist( f_{\theta}(X_0, X_1), f_{\theta}(X_0, P_1) ) = \\ JSD(s(f_{\theta}(X_0, X_1)), s(f_{\theta}(P_0, X_1)))
% \end{align*}
% \begin{align*}
%     \text{regularizer}_{\theta_0} = &+ \nabla JS(s(f_{\theta}(X_0, X_1)), s(f_{\theta}(X_0, P_1))) \\ &- \nabla JS(s(f_{\theta}(X_0, X_1)), s(f_{\theta}(P_0, X_1)))\\
%     \\
%     \text{regularizer}_{\theta_1} = &+ \nabla JS(s(f_{\theta}(X_0, X_1)), s(f_{\theta}(P_0, X_1))) \\ &- \nabla JS(s(f_{\theta}(X_0, X_1)), s(f_{\theta}(X_0, P_1)))\\
% \end{align*}

\newpage

{
    \small
    \bibliographystyle{ieeenat_fullname}
    \bibliography{main}
}

\end{document}